\documentclass[fleqn,usenatbib]{mnras}

\usepackage[T1]{fontenc}

\DeclareRobustCommand{\VAN}[3]{#2}
\let\VANthebibliography\thebibliography
\def\thebibliography{\DeclareRobustCommand{\VAN}[3]{##3}\VANthebibliography}

\usepackage{graphicx}	
\usepackage{booktabs}
\usepackage{hyperref}
\hypersetup{
    colorlinks=true,
    linkcolor=blue,
    filecolor=magenta,      
    urlcolor=cyan,
    pdftitle={Main},
    pdfpagemode=FullScreen,
    }
\usepackage{amsmath,amsfonts,amssymb}
\usepackage{color}
\usepackage{natbib}
\usepackage{epsfig}
\usepackage{float}
\usepackage{afterpage}
\usepackage{ifthen}
\usepackage[morefloats=12]{morefloats}
\usepackage{placeins}
\usepackage{multicol}
\usepackage{multirow}
\usepackage{siunitx}
\usepackage{pdflscape}
\DeclareSIUnit\parsec{pc}
\DeclareSIUnit\lightyear{ly}
\DeclareSIUnit\year{yr}

\def\$H_0${\rm H_0}

\def\LCDM{$\Lambda$-CDM}
\renewcommand*\vec[1]{\ensuremath{\boldsymbol{#1}}}

\title[BMA with application to the Hubble tension]{A convenient approach to characterizing model uncertainty with application to early dark energy solutions of the Hubble tension}

   \author[S. Paradiso et al.]{S. Paradiso,$^{1,2}$
          M. DiMarco,$^{2}$
          M. Chen,$^{3}$
          G. McGee,$^{3}$
          W.J. Percival,$^{1,2,4}$
          \\
          $^{1}$ Waterloo Centre for Astrophysics, University of Waterloo, Waterloo, ON N2L 3G1, Canada \\ 
          $^{2}$ Department of Physics and Astronomy, University of Waterloo, Waterloo, ON N2L 3G1, Canada \\
          $^{3}$ Department of Statistics and Actuarial Sciences, University of Waterloo, Waterloo, ON N2L 3G1, Canada \\
          $^{4}$ Perimeter Institute for Theoretical Physics, 31 Caroline St North, Waterloo, ON N2L 2Y5, Canada
          }
   \date{}

\pubyear{2023}

\begin{document}
\label{firstpage}
\pagerange{\pageref{firstpage}--\pageref{lastpage}}
\maketitle

\begin{abstract}
Despite increasingly precise observations and sophisticated theoretical models, 
the discrepancy between measurements of $H_0$ from the cosmic microwave background or from Baryon Acoustic Oscillations combined with Big-Bang Nucleosynthesis versus those from local distance ladder probes---commonly known as the "$H_0$ tension"---continues to perplex the scientific community.
To address this tension, Early Dark Energy (EDE) models have been proposed as alternatives to 
\LCDM, as they can
change the observed sound horizon and the inferred Hubble constant from measurements based on this.  
In this paper, we investigate the use of Bayesian Model Averaging (BMA) to evaluate EDE as a solution to the $H_0$ tension. BMA consists of assigning a prior to the model and deriving a posterior as for any other unknown parameter in a Bayesian analysis.
BMA can be computationally challenging in that one must approximate the joint posterior of both model and parameters. Here we present a computational strategy for BMA that exploits existing MCMC software and combines model-specific posteriors post-hoc. In application to a comprehensive analysis of cosmological datasets, we quantify the impact of EDE on the $H_0$ discrepancy.
We find an EDE model probability of $\sim 90\%$ whenever we include the $H_0$ measurement from Type Ia Supernovae in the analysis, whereas the other data show a strong preference for the standard cosmological model. 
We finally present constraints on common parameters marginalized over both cosmological models. For reasonable priors on models with and without EDE, the $H_0$ tension is reduced by at least 20\%.
\end{abstract}




\section{Introduction}
\label{sec:Intro}

The standard cosmological model, \LCDM, has been remarkably successful in explaining various cosmological phenomena. However, the persisting discrepancy in the determination of the Hubble constant, $H_0$, has challenged the robustness of this model. $H_0$, the present-day expansion rate of the Universe, plays a pivotal role in our understanding of cosmic evolution, the age of the Universe, and the formation of large-scale structures. It is typically inferred using one of two approaches: local distance ladder measurements or constraints as part of a cosmological model from more general observations such as those from the Cosmic Microwave Background (CMB) or from the combination of Baryon Acoustic Oscillation (BAO) and Big-Bang Nuclesynthesis (BBN) observations.

 Leveraging the CMB data, the Planck satellite mission, reported a value of $H_0$ close to $68$kms$^{-1}$Mpc$^{-1}$ for \LCDM\ cosmologies \citep{planck2018-VI}. This is supported by standard ruler observations of BAO from the Sloan Digital Sky Survey (SDSS) and BBN, which favour similar values for the same models \citep{eboss-cosmology}. For the CMB the absolute measurement at present day results from the photon temperature and black-body assumption, giving the physical photon density. Density ratios giving the CMB power spectrum then give the matter and baryon densities, with the angular scales of the projected peak positions finally leading to a constraint on $H_0$ through the Angular Diameter distance. For BAO+BBN, the physics is similar: BBN observations of the baryon-to-photon ratio, together with the present-day photon density from the CMB temperature, fix the present day size of the BAO ruler. The projected scale of the observed BAO then constrains $H_0$ in a similar way to the CMB constraints. In contrast, observations from local distance ladder studies, such as the Hubble Space Telescope (HST) observations of Cepheid variable stars, favour a higher value of $H_0\sim74$kms$^{-1}$Mpc$^{-1}$  \citep{Riess:2019cxk}. These observations can be considered a more direct route to measure the present-day expansion rate, using distance estimates for individual objects together with their redshifts to directly measure the expansion rate. The divergence in measurements, referred to as the $H_0$ tension, has remained persistent for a number of years now and, in some cases, has intensified as more precise data become available.

One proposal to resolve the $H_0$ tension is the introduction of Early Dark Energy (EDE) into the cosmological model. EDE posits the existence of a scalar field in the early Universe, leading to an enhancement of the expansion rate during the epoch of matter-radiation equality \citep{Karwal:2016,Poulin:2019}. This additional early acceleration can alleviate the $H_0$ tension by modifying the early-time cosmic expansion and the position of the acoustic feature in the radiation and matter power spectra. The EDE model has gained popularity due to its capacity to reconcile the Planck and HST measurements, but the extent to which the data support this model is unclear.

To address the challenges of model selection and parameter estimation within the context of the $H_0$ tension and EDE, we employ Bayesian Model Averaging (BMA) as a powerful statistical tool \citep{Parkinsons:2013,Trotta:2008}. By treating the model itself as another unknown in a Bayesian framework, it simultaneously accounts for uncertainty in parameter estimates as well as model choice. BMA offers a principled approach to statistically weight the evidence from various models, enabling a comprehensive analysis of the viability of EDE as a potential solution to the $H_0$ tension. 

Standard methods for making Bayesian inferences from cosmological data consider only a single model and then use a method such as Markov-Chain Monte-Carlo (MCMC) to sample from the posterior \citep{Hastings:1970,Tierney:1994} conditional on that model. Accounting for model uncertainty via BMA, however, requires the joint posterior for both models and parameters; this requires accounting for both the likelihood and the model prior, which is not readily available in model-conditional samples. We outline a method not previously applied in cosmology that collapses these model-conditional samples post-hoc, thus avoiding the need for specialized software for BMA. 

We use this method to apply Bayesian Model Averaging to scrutinize the implications of the EDE model on resolving the $H_0$ tension. We carefully consider a range of observational datasets, including CMB measurements, Type Ia Supernovae and Baryon Acoustic Oscillations, to constrain the relevant cosmological parameters. By employing BMA, we assess the credibility of the EDE model as a viable explanation for the $H_0$ tension, while simultaneously exploring the possible range of cosmological scenarios.

The organization of this paper is as follows: in Section 2 we present the theoretical framework of the EDE model and its underlying physical motivations. Section 3 details the Bayesian Model Averaging methodology employed in this analysis. In Section 4, we present our results and discuss the implications of the EDE model on the resolution of the $H_0$ tension. In Section 5 we conclude with a discussion of our findings and potential avenues for further exploration.

\section{Early Dark Energy}
\label{Sec:EDE}

The EDE model introduces an additional dark energy component, characterized by a fractional energy density parameter $\Omega_{\rm{EDE}}$ and associated equation of state (EOS) parameter $w_{\rm{EDE}}$, which is significant during the epoch of recombination. The presence of EDE modifies the cosmic expansion rate, leading to a different evolution of the scale factor $a(t)$ compared to the standard \LCDM\ model. During the epoch of recombination, the Friedmann equation for the Hubble parameter $H(t)$ can be written as
\begin{equation}
    H(t) = H_0\cdot\left( \Omega_{\rm{r}}a^{-4} + \Omega_{\rm{m}}a^{-3} + \Omega_\Lambda + \Omega_{\rm{EDE}}(a) \right)^{1/2}, 
    \label{eq:FRW}
\end{equation}
where $\Omega_{\rm{X}} \equiv \rho_{\rm{X}}/\rho_{\rm{crit}}$ is the density parameter for the X component (namely matter, radiation, $\Lambda$ and EDE). 
We assume that $\Lambda$ is constant and that the Universe is set up to have a flat geometry so that $\Omega_k=0$ and the curvature term does not appear in Eq.~\ref{eq:FRW}.

In this paper we focus on the theoretical EDE framework introduced by \citet{Smith:2020} based on a quintessence scalar field $\phi$ with an oscillating potential
\begin{equation}
    V_{\rm{n}}(\phi) = m^2f^2\left( 1 - \cos\left( \phi/f\right) \right)^n.
    \label{eq:EDE_field}
\end{equation}
The case $n=1$ represents the well established Axion potential. We will restrict our analysis to the case $n=3$, 
for the reasons given in \citet{Hill:2020}. The dynamics of the scalar field are governed by the Klein-Gordon equation
\begin{equation}
    \Ddot{\phi} + 3H\dot\phi + V_{\rm{n},\phi} = 0,
    \label{eq:EDE_KG}
\end{equation}
where $H$ is the Hubble parameter, the dots represent the derivative with respect to cosmic time and $ V_{\rm{n},\phi}=dV_{\rm{n}}/d\phi$.

\begin{figure}
    \centering
    \includegraphics[width=1\hsize]{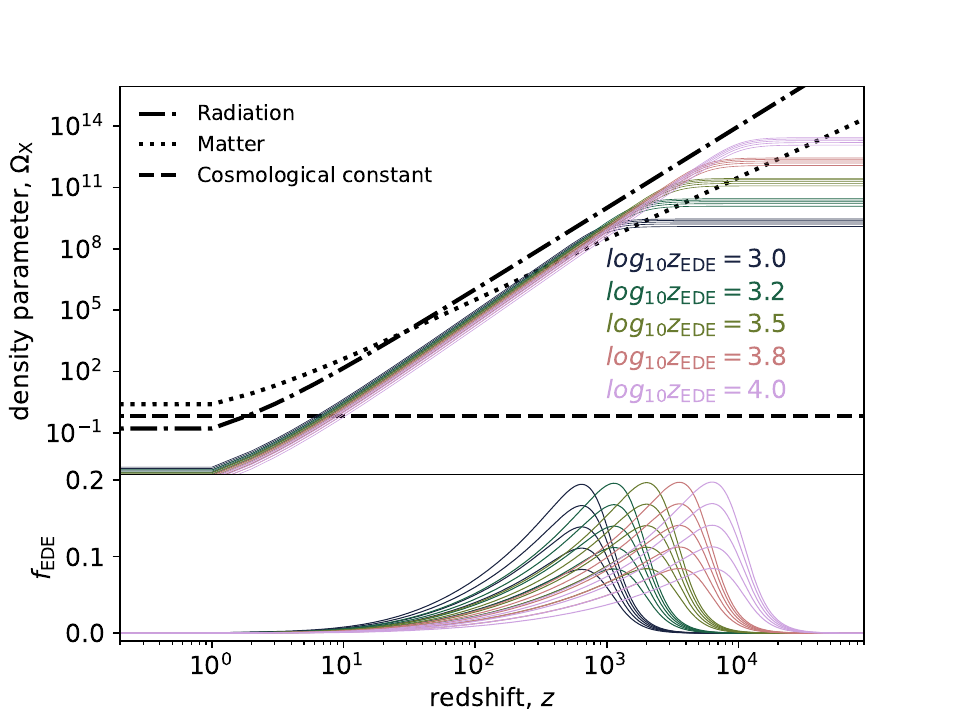}
    \caption{Redshift evolution of the density parameter $\Omega_{\rm X}$ for the main components of the Universe (top), and EDE fraction (bottom). We report the evolution of radiation (dot-dashed), matter (dotted), cosmological constant (dashed) and EDE (solid) spanning different combination of $z_{\rm{EDE}}$ in $[10^3,10^4]$ and $f_{\rm{EDE}}$ in $[0.6,1.4]$ with a darker to brighter color palette (for increasing values of $z_{\rm{EDE}}$). The redshift evolution of the EDE component is computed for $n=3$ in order to match the assumption in the main text. }
    \label{fig:EDE_plot}
\end{figure}

Heuristically speaking, in order to effectively mitigate the $H_0$ tension, the EDE contribution is concentrated in a short period after radiation-matter equality and before recombination, when the scalar field dominates the energy density contribution of the Universe and drives the growth of the scale factor before rapidly decaying. 
We refer the interested reader to \citet{Turner:1983}, \citet{Marsh:2010}, \citet{Poulin:2019}, and references therein, for the details.
It is useful to define a re-normalized field variable $\Theta=\phi/f$, so that Eq.~\ref{eq:EDE_KG} reads
\begin{equation}
    \Ddot{\Theta} + 3H\dot\Theta + \frac{1}{f^2}V_{\rm{n},\phi} = 0.
    \label{eq:KE_EDE_theta}
\end{equation}
The evolution of the scalar field can be completely determined by the set of three parameters $\{ m,f,\Theta_{\rm{i}}\}$, where $\Theta_{\rm{i}}$ represents the initial field value in units of $f$. From an observational point of view, we can characterize the background evolution by means of the maximum fraction of the total energy density in the field $f_{\rm{EDE}}$ and the redshift corresponding to that maximum $z_{\rm{c}}$, as heuristically illustrated in Fig.~\ref{fig:EDE_plot} for an axion-like EDE field \citep{Poulin:2018}; for a given combination of $\{m,f\}$ we can always find a value of $\{f_{\rm{EDE}},z_{\rm{c}}\}$, so that the model's free parameter set consists of either $\{ m,f,\Theta_{\rm{i}}\}$ or $\{ f_{\rm{EDE}},z_{\rm{c}},\Theta_{\rm{i}}\}$.

\section{Introduction to Bayesian Model Averaging}
\label{Sec:BMA}

In many statistical analyses, we are faced with the task of selecting the most appropriate model from a set of competing models. Traditional methods first fit a set of candidate models, and then compare their fit according to some criteria like the Akaike Information Criterion \citep[AIC]{Akaike:1974}; finally, inferences are drawn with respect to the ``best''-fitting model. A limitation of this approach is that inferences are typically drawn as if the model was chosen a priori---that is, it does not account for uncertainty in the choice of model. 

By contrast in a Bayesian framework, model uncertainty can be handled by treating the model like any other unknown, and computing the posterior probability of each candidate model given the observed data and chosen prior. This allows for a principled approach to comparing models and quantifying their relative plausibility based on the data and prior knowledge.

In addition to making direct inferences about the probabilities of different models, this approach allows one to formally account for model uncertainty when making inferences about cosmological parameters via Bayesian model averaging, or BMA. Intuitively, BMA computes a weighted average over the conditional posteriors for cosmological parameters under each candidate model. This marginalization occurs over the model posterior, ensuring that models with higher probabilities contribute more to the final inference.

Let $M=1,2,...,K$ be a discrete random variable indicating a choice among $K$ different models. For convenience, we will refer to $M=j$ as $M_j$. The posterior model probability for model $M_{\rm{i}}$, denoted as $P(M_{\rm{i}}\vert\Vec{d})$, is computed as
\begin{equation}
    P(M_{\rm{i}}\vert\Vec{d}) = \frac{ P(M_{\rm{i}}) P(\Vec{d}\vert M_{\rm{i}} )}{ \sum_{\rm{j}=1}^{\rm{K}} P(M_{\rm{j}}) P(\Vec{d}\vert M_{\rm{j}}) },
    \label{eq:BMA}
\end{equation}
where $P(M_{\rm{i}})$ is the prior probability of model $M_{\rm{i}}$, $P(\Vec{d}\vert M_{\rm{i}})$ is the marginal likelihood of the data $\Vec{d}$ given model $M_{\rm{i}}$, and the denominator serves as a normalization factor to ensure that the probabilities sum to one. We assume that the prior probabilities are separable between model selection and model parameters.

The advantages of BMA are:
\begin{enumerate}
    \item Accounting for Model Uncertainty: BMA provides a comprehensive approach to handle model uncertainty, acknowledging that multiple models may be plausible given the available data \citep{Hoeting:1999}.
    \item Robustness: BMA is less sensitive to the specific choice of a single "best" model, reducing the risk of overfitting or underfitting the data \citep{Hoeting:1999}.
    \item Improved Predictive Performance: By averaging predictions from multiple models, BMA often leads to more accurate predictions compared to relying on a single model \citep{Madigan:1994}.
\end{enumerate}

\subsection{\texttt{Fast-MPC} and application to a two model problem}
Without loss of generality, consider two candidate models, $M_1$ and $M_2$, parametrized by $\Vec{\theta}_1$ and $\Vec{\theta}_2$ respectively, and let $\Vec{\theta}$ be the union of $\Vec{\theta}_1$ and $\Vec{\theta}_2$. A full Bayesian analysis would characterize the joint posterior of $M_i$ and $\Vec{\theta}_i$ for $i$=1,2, from which we aim to: (i) obtain the posterior probability of each candidate model, $P(M_{\rm{i}}\vert\Vec{d})$; and (ii) obtain the marginal posteriors of the target parameters, $P(\Vec{\theta}\vert \Vec{d})$. The former allows us to assess the extent the evidence in favour of either candidate model; the latter allows our inferences about cosmological parameters to formally account for uncertainty in the choice of model.

Following \citet{Hastie:2012}, characterizing the full joint posterior (for model and parameters) can be done in one of two ways: \textit{across-model} approaches and \textit{within-model} approaches. The most well-known method in the former category is the reversible jump MCMC \citep[RJ-MCMC]{Green:1995}, a variant of the Metropolis–Hastings algorithm (known as Metropolis-Hastings-Green) that allows for between-model steps in which the dimension of the parameter vector changes to accommodate models of different sizes. This is a general approach that allows one to explore a large space of models, but it requires somewhat bespoke software implementation (see, for example \citealp{Karnesis:2023ras}), and achieving convergence can be challenging in practice. 

Instead, we adopt a \textit{within-model} approach that exploits our knowledge of $P(\Vec{\theta}\vert \Vec{d},M_1)$ and $P(\Vec{\theta}\vert \Vec{d},M_2)$ which are easily obtained from the usual model-conditional MCMC \citep{Hastie:2012,Green:2000} using standard software. The marginal posterior of $\Vec{\theta}$ can be decomposed as follows:
\begin{align}
    P(\Vec{\theta}\vert \Vec{d}) &= P(\Vec{\theta}, M_1\vert \Vec{d}) + P(\Vec{\theta}, M_2\vert \Vec{d}) \\
    &= P(\Vec{\theta}\vert \Vec{d},M_1)P(M_1\vert\Vec{d}) + P(\Vec{\theta}\vert \Vec{d},M_2)P(M_2\vert\Vec{d}).
    \label{eq:marginal_posterior_BMA}
\end{align} 
Since samples of $P(\Vec{\theta}\vert \Vec{d},M_1)$ and $P(\Vec{\theta}\vert \Vec{d},M_2)$ are readily available from standard analyses conditional on a single model, it then only remains to obtain the two model posteriors $P(M_{\rm{i}}\vert\Vec{d})$ defined in Eq.~\ref{eq:BMA}. 
This decomposition frames the model-averaged posterior $P(\Vec{\theta}|d)$ as weighted average of the model-conditional posteriors $P(\Vec{\theta}|\Vec{d},M_i)$ as in Eq. \ref{eq:marginal_posterior_BMA}. Note, however, that it is not sufficient to weight by the model-priors $P(M_i)$; rather the weights correspond to the model posteriors $P(M_i|\Vec{d})$---the data themselves inform how much weight is assigned to either model.

The model posteriors $P(M_{\rm{i}}\vert\Vec{d})$ in Eq.~\ref{eq:BMA} consist of both the model priors, $P(M_{\rm{i}})$, which are specified by the analyst, as well as the marginal likelihood:
\begin{align}
    P(\Vec{d}\vert M_{\rm{i}}) &= \int P(\Vec{\theta}_{\rm{i}},\Vec{d}\vert M_{\rm{i}})d\Vec{\theta}_{\rm{i}}\\
    &=  \int P(\Vec{d}\vert M_{\rm{i}},\Vec{\theta}_{\rm{i}})P(\Vec{\theta}_{\rm{i}}\vert M_{\rm{i}})d\Vec{\theta}_{\rm{i}}\,.
    \label{eq:marginal_params}
\end{align}
for \rm{i}=1,2. We obtain an estimate $\hat{P}(\Vec{d}\vert M_{\rm{i}})$ of  $P(\Vec{d}\vert M_{\rm{i}})$ based on the model-conditional MCMC chain, as described in Sect.~7 of \citet{Newton:1994}:
\begin{equation}  \label{eq:marginal_like}
    \hat{P}(\Vec{d}\vert M_{\rm{i}}) = \frac{\sum_{\rm{t}=1}^{\rm{N}_{\rm{i}}} P(\Vec{d}\vert M_{\rm{i}},\Vec{\theta}_{\rm{i}}^{(\rm{t})}) \omega(\Vec{\theta}_{\rm{i}}^{(\rm{t})})}{\sum_{\rm{t}=1}^{\rm{N}_{\rm{i}}}\omega(\Vec{\theta}_{\rm{i}}^{(\rm{t})})}\,,
\end{equation}
where $\Vec{\theta}_{\rm{i}}^{(\rm{t})}$ is the t-th sample in the i-th model-conditional MCMC chain with length $\rm{N}_{\rm{i}}$, and $\omega(\Vec{\theta}_{\rm{i}})=P(\Vec{\theta_{\rm{i}}}\vert M_{\rm{i}})/\phi(\Vec{\theta}_{\rm{i}})$. This is effectively importance sampling to perform a Monte-Carlo integration over the prior $P(\Vec{\theta_{\rm{i}}}\vert M_{\rm{i}})$ with an importance sampling function $\phi(\Vec{\theta}_{\rm{i}})$, which is set to the model-conditional posterior $P(\Vec{\theta}_{\rm{i}}\vert M_{\rm{i}},\Vec{d})$ so that we can use the readily available model-conditional chains.

To see this, by Bayes' theorem, we can express Eq.~\ref{eq:marginal_params} as
\begin{align}
    P(\Vec{d}\vert M_{\rm{i}}) &= \frac{P(\Vec{d}, \Vec{\theta}_{\rm{i}} \vert M_{\rm{i}})}{P(\Vec{\theta}_{\rm{i}}\vert\vec M_{\rm{i}},\Vec{d})} \\
    &=\frac{P(\Vec{\theta}_{\rm{i}}\vert M_{\rm{i}})P(\Vec{d}\vert\Vec{\theta}_{\rm{i}}, M_{\rm{i}})}{P(\Vec{\theta}_{\rm{i}}\vert M_{\rm{i}},\Vec{d})} = \omega(\Vec{\theta}_{\rm{i}})P(\Vec{d}\vert M_{\rm{i}},\Vec{\theta}_{\rm{i}})\,.
    \label{eq:marginal_params_var}
\end{align}
where $P(\Vec{d}\vert M_{\rm{i}},\Vec{\theta}_{\rm{i}})$ is the likelihood conditional on model $M_i$. From this we have
$\omega(\Vec{\theta}_{\rm{i}})=P(\Vec{d}\vert M_{\rm{i}})/P(\Vec{d}\vert M_{\rm{i}},\Vec{\theta}_{\rm{i}})$. Substituting this into Eq.~\ref{eq:marginal_like}, we see that the estimator is simply the harmonic mean of the likelihood, averaging over the model-conditional  chain
\begin{equation} 
    \hat{P}(\Vec{d}\vert M_{\rm{i}}) =\frac{\rm{N}_{\rm{i}}}{\sum_{\rm{t}=1}^{\rm{N}_{\rm{i}}}\left[ 1/P(\Vec{d}\vert M_{\rm{i}}, \Vec{\theta}_{\rm{i}}^{(\rm{t})})\right]}\,.
    \label{eq:BMA_like_estimator}
\end{equation}
To compute this, one must have access to (i) the samples $\theta_{\rm{i}}^{(\rm{t})}$ from the model conditional MCMC,  as well as (ii) the full likelihoods evaluated at each of those samples $P(\Vec{d}\vert M_{\rm{i}}, \Vec{\theta}_{\rm{i}}^{(\rm{t})})$, both of which are easily accessible using a cosmological MCMC code such as \texttt{Cobaya}\ \citep{Torrado:2020dgo,Lewis:2002ah,Lewis:2013hha}. It is shown in \citet{Newton:1994} that this estimator converges almost surely to the true $P(\Vec{d}\vert M_{\rm{i}})$. From this and the model priors we can estimate $P(M_{\rm{i}}\vert\Vec{d})$ via Eq.~\ref{eq:BMA}.

We have implemented this method in a publicly available software package called \texttt{Fast-MPC} available at \hyperlink{https://github.com/simonpara/Fast-MPC}{https://github.com/simonpara/Fast-MPC}. \texttt{Fast-MPC} implementation facilitates BMA via the following steps:
\begin{enumerate}
    \item Fit candidate models: run two MCMC chains assuming $M_1$ and $M_2$ respectively, yielding $\{\Vec{\theta}_1\}_{\rm{N}_1}$, $\{\Vec{\theta}_2\}_{\rm{N}_2}$, $\{P(\Vec{d}\vert M_{\rm{1}}, \Vec{\theta}_{\rm{1}})\}_{\rm{N}_1}$ and $\{P(\Vec{d}\vert M_{\rm{2}}, \Vec{\theta}_{\rm{2}})\}_{\rm{N}_2}$.
    \item Estimate ${P}(\Vec{d}\vert M_{1})$ and ${P}(\Vec{d}\vert M_{2})$: compute estimates $\hat{P}(\Vec{d}\vert M_{1})$ and $\hat{P}(\Vec{d}\vert M_{2})$ via Eq.~\ref{eq:BMA_like_estimator}.
    \item Estimate model posteriors: substitute estimates $\hat{P}(\Vec{d}\vert M_{1})$ and $\hat{P}(\Vec{d}\vert M_{2})$ into Eq.~\ref{eq:BMA} to estimate $P(M_i|d)$.
    \item Estimate marginal (model-averaged) posteriors of $\Vec{\theta}$: substitute estimates of $P(M_i|d)$ into Eq.~\ref{eq:marginal_posterior_BMA}.
\end{enumerate}

Alternatively, reversible-jump MCMC (RJ-MCMC)  would instead  produce a single MCMC chain spanning multiple models (An example implementation is \texttt{Eryn}\footnote{\hyperlink{https://github.com/mikekatz04/Eryn.git}{https://github.com/mikekatz04/Eryn.git}} described in \citealt{Karnesis:2023ras}). The resulting chain could easily be used to determine model probabilities as for any other parameter, and would not require us to compute a direct estimate of $P(\Vec{d}\vert M_{\rm{i}})$. In practice, however, convergence is a challenge using RJ-MCMC, and \texttt{Fast-MPC} allows for a faster and more convenient computation when (1) the number of candidate models is small and (2) a MCMC sample for each model is available (be it from off-the-shelf software or from publicly available data releases). We compare the two approaches in Appendix~\ref{appendix:posterior_weighting}, validating that they give consistent answers in a  toy example.

\section{Results}
\label{Sec:Results}

For our analysis, we adopted the publicly available MCMC sampler for cosmology \texttt{Cobaya}\footnote{\hyperlink{https://github.com/CobayaSampler/cobaya}{https://github.com/CobayaSampler/cobaya}} \citep{Torrado:2020dgo,Lewis:2002ah,Lewis:2013hha}; we exploited theory codes \texttt{CAMB} \citep{Lewis:1999bs,Howlett:2012mh} and its Early Quintessence model implementation \citep{Smith:2020}. The BMA was performed through \texttt{Fast-MPC} with our publicly available code\footnote{\hyperlink{https://github.com/simonpara/Fast-MPC}{https://github.com/simonpara/Fast-MPC}}; it consists of a library of functions for reading-in MCMC samples and computing all the necessary quantities to estimate $P(M_{\rm{i}}\vert\Vec{d})$.

Our datasets included the following observables from CMB, CMB lensing, BAO and SNIa:
\begin{itemize}
    \item \textbf{CMB}: Planck 2018 CMB TT, TE and EE likelihood \citep{planck2018-V} in the high-$\ell$ multipole range $30\le\ell\le 2508$ (TT) and $30\le\ell\le 1996$ (TE,EE), and the low-$\ell$ \texttt{Commander} TT likelihood spanning $2\le\ell\le 29$.
    \item \textbf{Lensing}: Planck 2018 reconstructed CMB lensing potential power spectrum \citep{planck2018-VIII}.
    \item \textbf{BAO}: Baryonic Acoustic Oscillation measurements from SDSS DR7 main galaxy sample \citep{Ross:2014qpa} at $z=0.15$, the 6dF galaxy redshift survey \citep{Beutler:2012px} at $z=0.106$ and the BOSS DR12 \citep{Alam:2016hwk} LOWZ and CMASS galaxy samples at $z=0.38$, $z=0.51$ and $z=0.61$, the eBOSS galaxies and quasars in $0.6\le z\le 2.2$ and the Lyman-$\alpha$ forest samples in $1.8\le z\le 3.5$.
    \item \textbf{SNIa}: 1048 luminosity distance measurements from the Pantheon SNIa light curves catalogue \citep{Scolnic:2017caz} and Cepheid measurements in the Large Magellanic Cloud from \citet{Riess:2019cxk}.
\end{itemize}

We considered the following dataset combinations: CMB, CMB+Lensing+BAO+SN, BAO+BBN and BAO+BBN+SN. The Planck-free cases include information from BBN in the form of a Gaussian prior of $\Omega_bh^2=0.02235\pm0.00037$ under the assumption of $N_{\rm{eff}}=3.046$. This prior is derived using the empirically-estimated cross-section of the deuterium described in \citet{Aldaberger:2011}, with abundance $\mathrm{D}/\mathrm{H} = (2.527\pm0.030)\times 10^{-5}$ from the high-resolution spectroscopic measurements of seven quasar absorption systems \citep{Cooke:2018}.

\begin{figure}
    \centering
    \includegraphics[width=1\hsize]{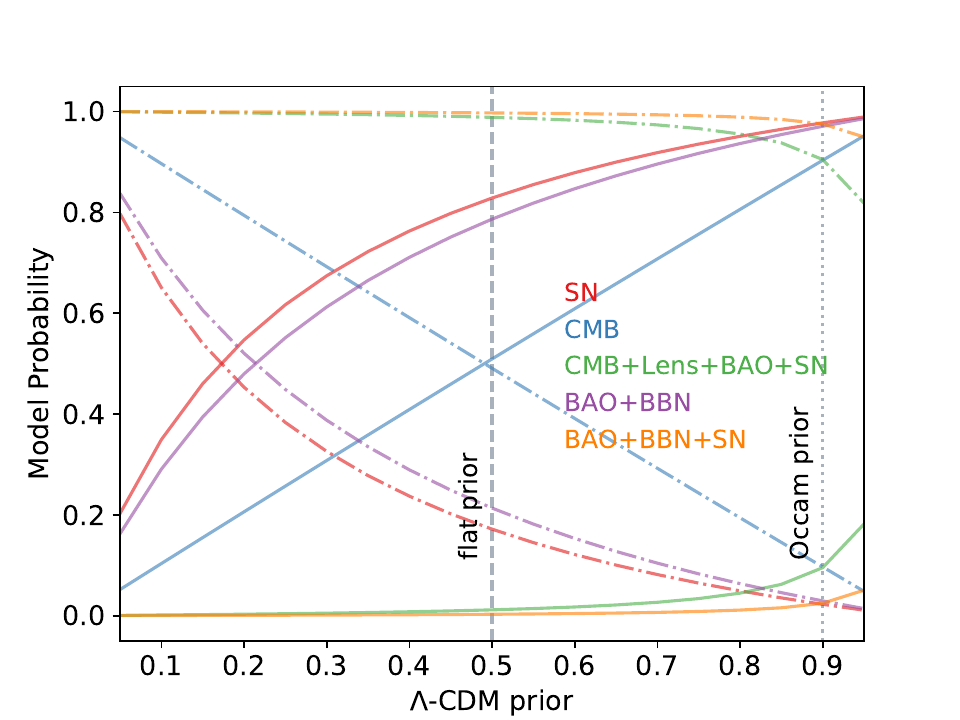}
    \caption{Model prior dependence of model posteriors. The models being compared are the $\Lambda$-CDM model (solid lines) and the EDE model (dot-dashed lines) for the considered datasets: SN (red), CMB (blue), BAO+BBN (green), CMB+Lensing+BAO+SN (purple), BAO+BBN+SN (orange). We reported as a grey vertical line the flat-prior (dashed) and the Occam's prior (dotted).}
    \label{fig:modelpriordep}
\end{figure}

As introduced in Sect.~\ref{Sec:BMA}, BMA allows us to compute the posterior probability of our model, given the data and the cosmological parameters, in a Bayesian way, and therefore relies on a prior assumption on model probability as shown in Eq.~\ref{eq:BMA}. We show in Fig.~\ref{fig:modelpriordep} the computed values of $P(M\vert\Vec{d})$ as a function of the prior assumption on the \LCDM\ model (or, equivalently, $1-P(M_{\rm{EDE}})$), after removing a conservative burn-in fraction of $0.5$ . The solid lines are the \LCDM\ model probabilities and the dot-dashed lines are the corresponding EDE model probabilities for the different considered datasets. We identified two interesting prior configurations, represented by the grey dotted and dashed lines on the plot:
\begin{itemize}
    \item \textbf{the flat-prior}, corresponding to a 50-50 split probability between \LCDM\ and EDE, and reflecting the agnostic assumption on which model should be favoured. This is the default prior assumption throughout the paper;
    \item \textbf{the Occam's prior}, expressing an a-priori preference for the \LCDM\ model, as it comes with fewer parameters. Note that some observations (such as supernovae samples) are cumulative, or the sample variance is the same (such as the CMB), and hence it is not good practice to form a prior based on data that are also used in the likelihood. We therefore do not include in this prior any previous match between observations and the \LCDM\ model, but we include results using this prior as a counter to the flat-prior, allowing us to test the sensitivity to the prior. We choose a 90\% probability in favour of the \LCDM\ cosmological model. As well as explicitly putting in a preference for a simpler model as we do here, we note that BMA also generally favours models with fewer parameters, because they have a smaller volume over which the likelihood is marginalized. Thus the idea behind Occam's razor does not only arise in the prior in BMA. For this reason, this prior choice results in strong advocacy for the \LCDM\ model and its interpretation should be made carefully. However, the Occam's prior represents an extreme case supporting the \LCDM\ model, and we report results under this assumptions alongside with the standard flat-prior for completeness.
\end{itemize}
We report the model probabilities under both prior assumptions in Table~\ref{tab:model_posteriors_vs_priors}. The \LCDM\ model is favoured for both priors for BAO+BBN and SN datasets when considered alone, with a \LCDM\ probability of $79\%$ and $83\%$ respectively when we adopt a flat model-prior---\LCDM\ is favoured above $90\%$ under the Occam's prior. 
The CMB-only case model probability closely matches the prior probabilities with $51\%$ and $91\%$ \LCDM\ model probabilities in the flat-prior and Occam's prior respectively; a straightforward explanation for this is that the data are not informative about the model choice, and all the information is coming from the prior. 
On the other hand, we find a strong EDE preference whenever we merge the datasets into CMB+Lensing+BAO+SN and BAO+BBN+SN, with more than $99\%$ EDE model probability under a flat model-prior, and, respectively, $96\%$ and $98\%$ for the Occam's prior. It is not a surprise that the joint datasets BAO+BBN+SN and CMB+Lensing+BAO+SN show a strong preference for EDE; that was precisely the reason EDE was introduced in the first place. These numbers should not be interpreted as one model being right or wrong from an absolute perspective, but rather the probability of one or the other to be favoured by data. I.e. our averaged constraints are still conditioned on one of the two models ($\Lambda$CDM or EDE) being correct. 

A further goal of our study is to incorporate the model uncertainty in the final cosmological parameter estimates. If there are big differences between the credible intervals for a particular parameter conditioned on different models, then we might expect the credible interval to be wider when averaged over models. Thus BMA has the general ability to reduce tensions by providing wider credible intervals given the model uncertainty. 
This explains why, for the $H_0$ tension, the marginalized uncertainty tends to mitigate the tension on cosmological parameters, especially on $H_0$, as we find in the following section.

\begin{table}
    \centering
    \begin{tabular}{lcr}
    \hline\hline
    & \multicolumn{2}{c}{Model Posterior (\LCDM,EDE) in} \\
    & \multicolumn{2}{c}{$[ P(\rm{\Lambda CDM}\vert\Vec{d}),  P(\rm{EDE}\vert\Vec{d}) ]$ in \%} \\
    \cmidrule{2-3}
    Dataset                   & Flat-prior         & Occam's-prior   \\
    \hline
    CMB                       & $[ 50.9,49.1 ]$ & $[ 90.3,9.7 ]$  \\
    CMB+Lensing+BAO+SN        & $[ 1.2,98.8 ]$  & $[ 9.5,90.5 ]$  \\
    BAO+BBN                   & $[ 78.7,21.3 ]$ & $[ 97.1,2.9 ]$  \\
    BAO+BBN+SN                & $[ 0.3,99.7 ]$  & $[ 2.5,97.5 ]$  \\
    SN                        & $[ 82.8,17.2]$  & $[ 97.7,2.3 ]$  \\
    \hline
    \end{tabular}
    \caption{Model posteriors as $[ P(\rm{\Lambda CDM}\vert\Vec{d}),  P(\rm{EDE}\vert\Vec{d}) ]$ for two model prior assumptions: 1) the flat-prior is a $50\%$ prior for each model and 2) the Occam's prior with a $90\%$ prior on \LCDM\ versus a $10\%$ prior on EDE.}
    \label{tab:model_posteriors_vs_priors}
\end{table}

\subsection{Model-Averaged Posteriors}
The knowledge of $P(M\vert\Vec{d})$ allows us to compute the cosmological parameters' posterior distributions marginalized over the model uncertainty as described in Eq.~\ref{eq:marginal_posterior_BMA}. 
For our analysis, we considered the following combination of the \LCDM\ base parameters with their flat prior ranges, unless otherwise specified: today's value of the Hubble parameter $H_0\in[20,100]$, the baryon density parameter $\Omega_{\rm{b}}\rm{h}^2\in[0.005,0.1]$ (and a Gaussian prior $\mathcal{N}(0.00235,(0.00037)^2)$ when including BBN), the cold dark matter density parameter $\Omega_{\rm{c}}\rm{h}^2\in[0.001,0.99]$, the amplitude of scalar perturbations $\log(10^{10} A_\mathrm{s})\in[1.61,3.91]$, the index of the scalar perturbations $n_{\rm{s}}\in[0.8,1.2]$ and the optical depth of reionization $\tau\in[0.01,0.8]$. In addition, we sampled the EDE parameters already introduced in Sect.~\ref{Sec:EDE} by keeping $n=3$ fixed and adopting the following flat priors: $\Theta_{\rm{i}}\in[0.1,3.1]$, $f_{\rm{EDE}}\in[0.01,0.5]$ and $\log (10\; z_{\rm{EDE}})\in[3.2,4.2]$.

We report in Table~\ref{tab:LCDM_marginalized_constraints} the constraints for the six \LCDM\ base parameters after marginalizing over the model uncertainty with a flat model prior (top row for each parameter) and the Occam's model prior (bottom row). We report only the common parameters between the \LCDM\ and the EDE models, for those are the only ones influenced by the model uncertainty. For the reader's reference, we provide the EDE specific model parameters' constraints in Table~\ref{tab:EDE_constraints} as evaluated in our analysis. 

\begin{table*}
    \centering
    \begin{tabular}{lcccr}
    \hline\hline
    Parameter & CMB & CMB+Lensing+BAO+SN & BAO+BBN & BAO+BBN+SN \\
    \hline\\[-2ex]
    \multirow{2}*{$H_0$}                      & $68.17^{+0.77}_{-1.4}$          & $71.1\pm 1.0$                  & $70.3^{+1.4}_{-2.7}$        & $74.0\pm 1.4$        \\[0.5ex]
                                              & $67.45^{+0.71}_{-0.89}$         & $71.0^{+1.4}_{-1.0}$           & $68.8^{+1.6}_{-2.4}$        & $73.9\pm 1.5$        \\[0.5ex]
                                              \hline\\[-2ex]
    \multirow{2}*{$\Omega_bh^2$}              & $0.02250^{+0.00017}_{-0.00025}$ & $0.02284\pm 0.00022$           & $0.02236\pm 0.00037$        & $0.02236\pm 0.00037$ \\[0.5ex]
                                              & $0.02239^{+0.00014}_{-0.00018}$ & $0.02283\pm 0.00023$           & $0.02236\pm 0.00037$        & $0.02236\pm 0.00037$ \\[0.5ex]
                                              \hline\\[-2ex]
    \multirow{2}*{$\Omega_ch^2$}              & $0.1229^{+0.0019}_{-0.0042}$    & $0.1291\pm 0.0039$             & $0.1240^{+0.0098}_{-0.013}$ & $0.1398\pm 0.0089$   \\[0.5ex]
                                              & $0.1207^{+0.0011}_{-0.0021}$    & $0.1284^{+0.0051}_{-0.0038}$   & $0.118^{+0.011}_{-0.013}$   & $0.1397\pm 0.0090$   \\[0.5ex]
                                              \hline\\[-2ex]
    \multirow{2}*{$\log (10^{10}A_{\rm{s}})$} & $3.050\pm 0.017$                & $3.065\pm 0.015$               & $3.12\pm 0.13$              & $3.08\pm 0.11$       \\[0.5ex]
                                              & $3.046\pm 0.017$                & $3.064\pm 0.015$               & $3.15\pm 0.13$              & $3.08\pm 0.11$       \\[0.5ex]
                                              \hline\\[-2ex]
    \multirow{2}*{$n_{\rm{s}}$}               & $0.9700^{+0.0052}_{-0.0089}$    & $0.9865\pm 0.0070$             & $-$                         & $-$                  \\[0.5ex]
                                              & $0.9658^{+0.0042}_{-0.0056}$    & $0.9854^{+0.0085}_{-0.0071}$   & $-$                         & $-$                  \\[0.5ex]
                                              \hline\\[-2ex]
    \multirow{2}*{$\tau$}                     & $0.0549\pm 0.0080$              & $0.0579\pm 0.0072$             & $-$                         & $-$                  \\[0.5ex]
                                              & $0.0544\pm 0.0079$              & $0.0580\pm 0.0073$             & $-$                         & $-$                  \\[0.5ex]
    \hline
    \end{tabular}
    \caption{Constraints on cosmological parameters marginalized over the model uncertainty with the flat model prior assumption (top row) and the Occam's prior (bottom row). We report $68\%$ credible intervals for each parameter.} 
    \label{tab:LCDM_marginalized_constraints}
\end{table*}

\begingroup
\setlength{\tabcolsep}{3pt} 
\begin{table}
    \centering
    \begin{tabular}{cccr}
    \hline\hline
    Dataset  & $\Theta_{\rm{i}}$ & $f_{\rm{EDE}}$ & $\log (10\; z_{\rm{EDE}})$ \\
    \hline\\[-2.5ex]
    CMB                                                                          & $2.15^{+0.93}_{-0.31}$                  & $0.045^{+0.010}_{-0.034}$                & $3.65\pm 0.21$                          \\[0.3ex]
    \multirow{2}*{\centering\begin{tabular}{c}CMB+Lensing\\+BAO+SN\end{tabular}} & \multirow{2}*{$2.58^{+0.35}_{+0.036}$}  & \multirow{2}*{$0.101^{+0.035}_{-0.029}$} & \multirow{2}*{$3.62^{+0.22}_{-0.15}$}   \\\\[0.3ex]
    BAO+BBN                                                                      & $2.601^{+0.057}_{-0.17}$                & $0.110^{+0.077}_{-0.10}$                 & $3.573\pm 0.085$                        \\[0.3ex]
    \multirow{2}*{\centering\begin{tabular}{c}BAO+BBN\\+SN\end{tabular}}         & \multirow{2}*{$< 2.20$}                 & \multirow{2}*{$0.238^{+0.062}_{-0.083}$} & \multirow{2}*{$3.73\pm 0.29$}           \\\\
    \hline
    \end{tabular}
    \caption{Constraints on the EDE model parameters from the different datasets taken into consideration in our analysis.}
    \label{tab:EDE_constraints}
\end{table}
\endgroup

\begin{figure*}
    \centering
    \includegraphics[width=1\hsize]{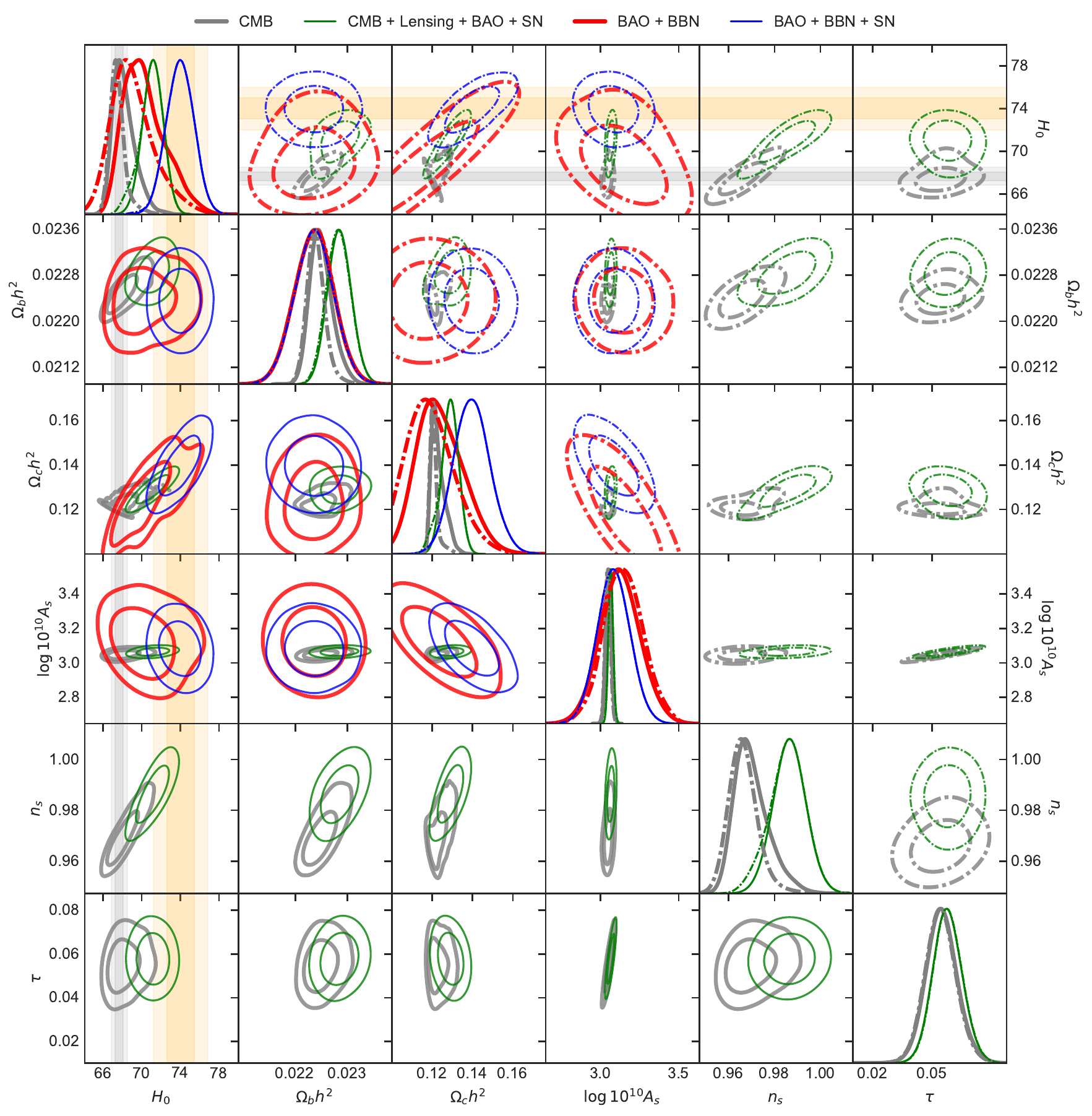}
    \caption{Comparison of the six \LCDM\ parameters marginalized over the model uncertainty with a flat model prior (bottom triangle, solid lines) and the Occam's prior (upper triangle, dot-dashed lines) from CMB (grey), CMB+Lensing+BAO+SN (green), BAO+BBN (red), and BAO+BBN+SN (blue); the parameters shown are the Hubble parameter today $H_0$, the baryon density parameter $\Omega_bh^2$, the cold dark matter density parameter $\Omega_ch^2$, the scalar perturbation amplitude $\log (10^{10}A_{\rm{s}})$, the scalar perturbation spectral index $n_{\rm{s}}$ and the optical depth of reionization $\tau$. For the BAO+BBN and BAO+BBN+SN, the last two parameters are not constrained by the data, and therefore not available in the plot. We included as vertical (horizontal) bands the the constraint (68\% and 95\% CL) on $H_0$ from Planck 2018 \citep{planck2018-VI} (gray), and from SNIa \citep{Riess:2019cxk} (orange), both under the \LCDM\ model assumption.}
    \label{fig:6pars_marg}
\end{figure*}

Fig.~\ref{fig:6pars_marg} shows the full posterior probability distributions, after the model uncertainty marginalization, for both the flat and the Occam's model priors. We shall make a few observations about the impact of the BMA on cosmological parameters' posteriors.
Firstly, the error bars increase with respect to the model-conditional parameter posterior error bars, and it is completely due to the additional model uncertainty considered in the analysis; the impact is more relevant with a flat model prior, especially for those datasets with a lower preference for one specific model, namely CMB and BAO+BBN. 
Because the EDE model was introduced to solve the Hubble tension, any preference for it using both SN and high-redshift data cannot be interpreted at the same level as if this was a model suggested before the tension became apparent. Ideally, we would therefore like to see a preference for the EDE model from one of other data set, rather than the combination. The results in Fig.~\ref{fig:6pars_marg} show that this is not the case, and the high-redshift data, when analysed on their own do not show a preference for either model.
The effect of the Occam's prior is mostly visible in those very datasets, but the effect is to shrink the parameters' uncertainties towards those of the model-conditional's parameter posteriors. Conversely, we notice the opposite behaviour for CMB+Lensing+BAO+SN and BAO+BBN+SN, where the model posterior expresses a strong preference for the EDE model with a flat model prior, whereas the Occam's prior increases the \LCDM\ model probability and therefore broadens the cosmological parameters' constraints. In the upper-left tile, and throughout the first column/row of the plot, we included the Planck 2018 68\% and 95\% confidence levels on $H_0$: from \citet{planck2018-VI} we report the TT+EE+TE+lowE+Lensing+BAO constraint of $H_0=67.66\pm0.42$, and similarly for SNIa \citep{Riess:2019cxk} we report the value $H_0=74.03\pm1.42$. Both these constraints assume a flat-\LCDM\ model, and interestingly, the $H_0$ tension is mitigated when accounting for the additional model uncertainty for most of the dataset combinations, from $\approx 1\sigma$ in the Planck-only case to $\lesssim 1\sigma$ in the BAO+BBN case, with a flat model prior.
This does not come as a surprise. In fact, we would ideally expect such posterior incompatibility to be even further reduced by accounting for more models; those may include new and different physics or different data modelling.

\begin{figure}
    \centering
    \includegraphics[width=\hsize]{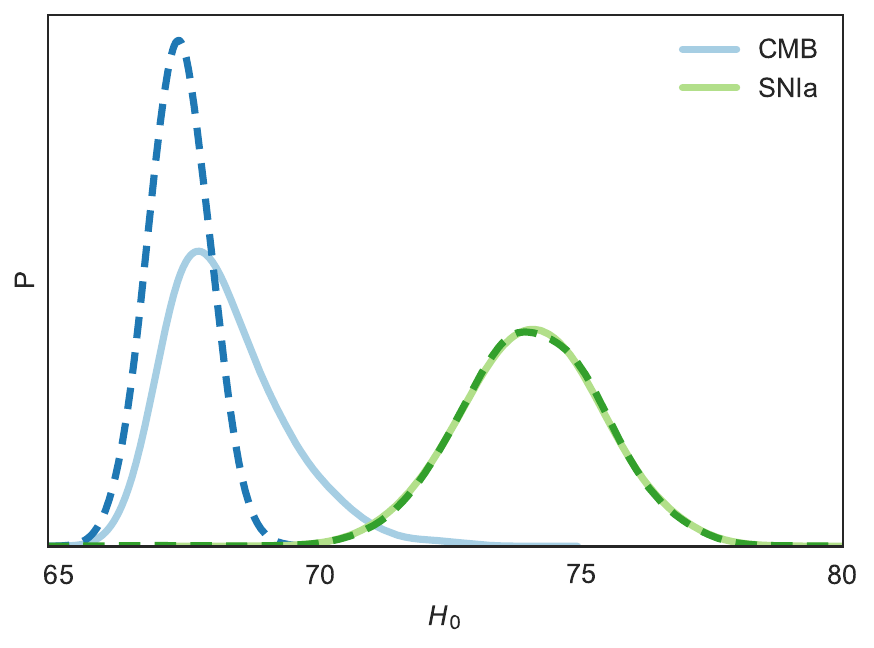}
    \caption{Posterior distribution of $H_0$ from CMB (blue) and SNIa (green) under the \LCDM\ model assumption (dashed line) and after marginalizing over the model uncertainty between \LCDM\ and EDE (solid line), assuming a flat model prior.}
    \label{fig:H0concordance}
\end{figure}

In the context of the $H_0$ tension, we present in Fig.~\ref{fig:H0concordance} a direct comparison of $H_0$ from CMB and SNIa datasets independently. This is the very original argument leading to the notorious $H_0$ tension. The dashed lines refer to the constraints obtained assuming (conditioning on, in the probabilistic context) the \LCDM\ model. Solid lines show the constraints after marginalizing over the model uncertainty by taking \LCDM\ and EDE into account. The $H_0$ posterior distribution from CMB broadens, and the parameter estimate changes from $67.29\pm0.61$ in the \LCDM\ model to $68.17^{+0.77}_{-1.4}$ after the marginalization, with a relative increase in the standard deviation of
\begin{equation}
    \Delta \sigma(H_0) = \frac{\sigma(H_0)^{\rm{marginal}} - \sigma(H_0)^{\Lambda\rm{CDM}}}{\sigma(H_0)^{\Lambda\rm{CDM}}} \approx 92\%,
\end{equation}
relieving the $H_0$ tension by $\approx 1\sigma$. The $H_0$ constraint from SNIa, on the other hand, shows no difference in the final posterior between the conditional and the model marginal case. This is due to the very high \LCDM\ model probability of the SN data that is mostly attributed to the insensitivity of SNIa measurements to the EDE parameters. Another way to understand this is that the model selection favours \LCDM\ by virtue of the smaller number of model parameters necessary to describe the SNIa data. 

\subsection{Assessing Convergence}
With any MCMC strategy, it is important to assess convergence in order to ensure one is sampling from the appropriate target distribution (the posterior). In the RJ-MCMC approach, where the model itself is sampled just like any other parameter, one must also consider convergence of the model choice. By contrast in the within-model strategy, the model is not sampled as part of the MCMC; rather, one would assess convergence for cosmological parameters separately for each model-conditional MCMC according to the Gelman-Rubin criterion on $R-1$ \citep{gelman:1992}. Once these chains have converged, however, a natural question is whether one has achieved a stable estimate of the model posterior.  

Consider an analogous Gelman-Rubin statistic for the collapsed chains:
\begin{equation}
    R = \sqrt{\frac{\sigma^2_{\rm{BC}}+\sigma^2_{\rm{WC}}}{\sigma^2_{\rm{WC}}}},
    \label{eq:R}
\end{equation}
where $\sigma^2_{\rm{BW}}$\footnote{It follows the usual definition: $\sigma^2_{\rm{BW}}=\frac{1}{\rm{J}-1}\sum_{\rm{i}=1}^{\rm{J}}(\bar{x}_{\rm{i}}-\langle\bar{x}_{\rm{i}}\rangle)^2$, where $\bar{x_{\rm{i}}}$ is the parameter's average for the i-th chain, J is the number of chains, and $\langle\bar{x}_{\rm{i}}\rangle$ is the average between all the chains.} and $\sigma^2_{\rm{WC}}$ refer to the between-chain variance and the within-chain variance of the model posterior probability $P(M_{\rm{i}}\vert\Vec{d})$, respectively. The latter is actually to be intended, in our context, as the collapsed within-chain variance if we stack the i-th chains from each MCMC conditional run in order to compute the model posterior probability. 
Moreover, there is a further complication with the evaluation of the within-chain variance of the model probability. Because Eq.~\ref{eq:BMA_like_estimator} is based on the harmonic mean, to estimate the within chain variance, we use a second-order Taylor expansion of $P(M_{\rm{i}}\vert\Vec{d})$ described in Appendix~\ref{appendix:wc-variance}.

\begin{figure}
    \centering
    \includegraphics[width=1\hsize]{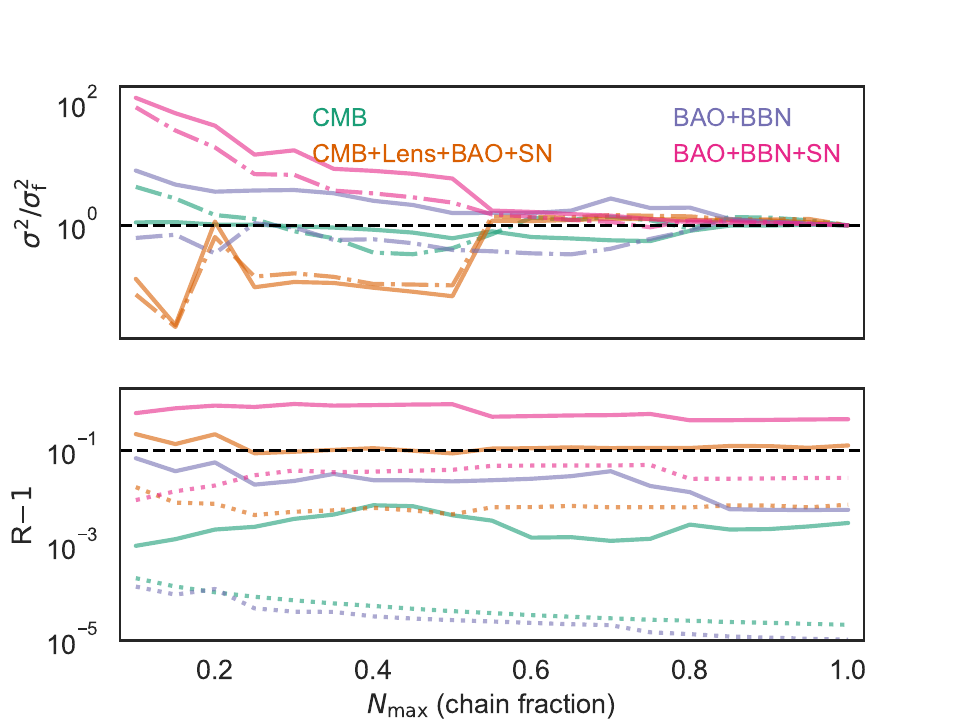}
    \caption{In the bottom panel, R-1 for the considered datasets as a function of the fraction of the overall number of samples. The solid line refer to the flat model prior, the dotted line the Occam's prior; we plot $R-1=0.1$ with a black dashed lines . On the top panel, the between-chains variance (solid line) and the within-chain variance (dashed) as a function of the overall number of samples, normalized by the value computed for the full chain length $\sigma^2_{\rm{f}}$. The black dashed lines refers to unitary value for the variance ratio.}
    \label{fig:gelman-rubin}
\end{figure}

The bottom panel of Fig.~\ref{fig:gelman-rubin} shows $R-1$ for the considered datasets as a function of the overall number of samples fraction. A good rule-of-thumb is to consider the chains \textit{converged} if $R-1<0.1$ (or equivalently $R<1.1$, \citealp{gelman:1992}). Our analysis shows that the CMB and BAO+BBN chains achieved the required convergence already with 10\% of samples under the flat-prior assumption, and the CMB+Lensing+BAO+SN with $\sim30\%$, whereas BAO+BBN+SN requires longer chains, $\sim 50\%$, to get R$-1\sim 0.1$. On the other hand, the Occam's prior assumption shows a more robust convergence, easily understood since we are casting the \LCDM\ model with a 90\% prior probability, and therefore most of the information is already in the model conditional chains for which convergence has already been assessed and achieved.

The top panel of Fig.~\ref{fig:gelman-rubin} shows the individual variance contributions to the computation of R, namely the between-chain variance (solid lines) and the collapsed within-chain variance (dashed lines)---referring to collapsed couples of model-conditional chains---as functions of the fraction of the total chain length and normalized for the value at the full chain length. This illustration can be very helpful to understand the behaviour of the variance of the model posterior distribution which, as already mentioned in this section, does not necessarily converge along with the within-model parameters' posteriors. In fact, we notice that it becomes stable for an overall chain length fraction of $\gtrsim 0.5-0.6$, for which the within-model parameters are still not converged.

\section{Conclusions}
\label{Sec:Conclusions}
We have presented an efficient BMA algorithm tailored for cosmological analyses called \texttt{Fast-MPC}. We have used this to investigate the pressing issue of the Hubble Constant tension ($H_0$ tension) within the framework of modern cosmology. Our work has demonstrated the power of BMA to decide between models in a clear and easy to understand way, while providing a robust and accurate characterization of the Universe's fundamental properties, by providing parameter constraints that marginalize over multiple models.

Through the application of \texttt{Fast-MPC} to a diverse set of cosmological datasets, including CMB, BAO, and Type Ia supernovae observations, we have 
considered the EDE solution to the Hubble tension in the Bayesian framework. Any Bayesian comparison between models must include a prior, as with any other parameter, and BMA makes these priors explicit.
By adopting a Bayesian perspective that incorporates model uncertainty, we have performed a more general cosmological parameter estimation, ultimately yielding more general constraints on cosmological parameters that allow EDE as a possible, but not the only, solution. We found that we cannot distinguish between the two models using CMB data alone, and therefore the additional model uncertainty leads to a $\sim 92\%$ larger credible interval for $H_0$ when a flat model prior is applied. On the other hand, BAO+BBN and SNIa datasets show a preference for the \LCDM\ model, with model uncertainty contributions to the overall $H_0$ posterior width of $\sim 155\%$ for the former, and null for the latter, for a flat model prior. Similarly, we found that the combined datasets CMB+lensing+BAO+SN and BAO+BBN+SN show a $\gtrsim 90\%$ EDE model probability for both flat and Occam's prior, with an uncertainty increase on $H_0$ of $\sim 210\%$ for CMB+lensing+BAO+BBN and $\sim 84\%$ for BAO+BBN+SN.

Our findings underscore the broader potential of BMA as a valuable tool for cosmological investigations. This work highlights the necessity of considering a diverse range of cosmological models and the importance of quantifying the associated uncertainties. Though the potential for BMA in cosmology has been acknowledged elsewhere (see \citealp{Parkinsons:2013}), uptake has been slow, possibly due to the computational challenges of between-model sampling strategies like RJ-MCMC. By contrast the simple computational approach presented here---which involves only post-hoc computation of standard model-conditional MCMC chains---has the potential to make BMA accessible in a wide array of problems, including large scale cosmological analyses.

In an era of precision cosmology where ever-more precise measurements are continually pushing the boundaries of our understanding, the \texttt{Fast-MPC} algorithm offers a robust and adaptable approach for model selection, accommodating the complexity of modern cosmological datasets. As we continue to refine our understanding of the Universe, BMA stands ready to contribute to more comprehensive and insightful cosmological investigations in the future.

\section*{Acknowledgements}

All authors acknowledge support from the Canadian Government through a New Frontiers in Research Fund (NFRF) Exploration grant.

WP acknowledges the support of the Natural Sciences and Engineering Research Council of Canada (NSERC), [funding reference number RGPIN-2019-03908] and from the Canadian Space Agency.

GM acknowledges the support of the Natural Sciences and Engineering Research Council of Canada (NSERC), [RGPIN-2022-03068 and DGECR-2022-004433].

Research at Perimeter Institute is supported in part by the Government of Canada through the Department of Innovation, Science and Economic Development Canada and by the Province of Ontario through the Ministry of Colleges and Universities.

This research was enabled in part by support provided by Compute Ontario (\hyperlink{computeontario.ca}{computeontario}) and the Digital Research Alliance of Canada (\hyperlink{alliancecan.ca}{alliancecan}).

\section*{Data Availability Statement}
The \texttt{Fast}-MPC code along with the plotting script used to generate the figures in this paper are publicly available at \hyperlink{https://github.com/simonpara/Fast-MPC}{https://github.com/simonpara/Fast-MPC}. All the data we used in this work are delivered within the MCMC sampler \texttt{cobaya}, at \hyperlink{https://github.com/CobayaSampler/cobaya}{https://github.com/CobayaSampler/cobaya}.



\bibliographystyle{mnras}
\bibliography{biblio} 

\begin{thebibliography}{}
\makeatletter
\relax
\def\mn@urlcharsother{\let\do\@makeother \do\$\do\&\do\#\do\^\do\_\do\%\do\~}
\def\mn@doi{\begingroup\mn@urlcharsother \@ifnextchar [ {\mn@doi@}
  {\mn@doi@[]}}
\def\mn@doi@[#1]#2{\def\@tempa{#1}\ifx\@tempa\@empty \href
  {http://dx.doi.org/#2} {doi:#2}\else \href {http://dx.doi.org/#2} {#1}\fi
  \endgroup}
\def\mn@eprint#1#2{\mn@eprint@#1:#2::\@nil}
\def\mn@eprint@arXiv#1{\href {http://arxiv.org/abs/#1} {{\tt arXiv:#1}}}
\def\mn@eprint@dblp#1{\href {http://dblp.uni-trier.de/rec/bibtex/#1.xml}
  {dblp:#1}}
\def\mn@eprint@#1:#2:#3:#4\@nil{\def\@tempa {#1}\def\@tempb {#2}\def\@tempc
  {#3}\ifx \@tempc \@empty \let \@tempc \@tempb \let \@tempb \@tempa \fi \ifx
  \@tempb \@empty \def\@tempb {arXiv}\fi \@ifundefined
  {mn@eprint@\@tempb}{\@tempb:\@tempc}{\expandafter \expandafter \csname
  mn@eprint@\@tempb\endcsname \expandafter{\@tempc}}}

\bibitem[\protect\citeauthoryear{Adelberger et~al.,}{Adelberger
  et~al.}{2011}]{Aldaberger:2011}
Adelberger E.~G.,  et~al., 2011, \mn@doi [Rev. Mod. Phys.]
  {10.1103/RevModPhys.83.195}, 83, 195

\bibitem[\protect\citeauthoryear{Aghanim et~al.}{Aghanim
  et~al.}{2020}]{planck2018-VIII}
Aghanim N.,  et~al., 2020, \mn@doi [Astron. Astrophys.]
  {10.1051/0004-6361/201833886}, 641, A8

\bibitem[\protect\citeauthoryear{Akaike}{Akaike}{1974}]{Akaike:1974}
Akaike H.,  1974, \mn@doi [IEEE Transactions on Automatic Control]
  {10.1109/TAC.1974.1100705}, 19, 716

\bibitem[\protect\citeauthoryear{Alam et~al.}{Alam et~al.}{2017}]{Alam:2016hwk}
Alam S.,  et~al., 2017, \mn@doi [Mon. Not. Roy. Astron. Soc.]
  {10.1093/mnras/stx721}, 470, 2617

\bibitem[\protect\citeauthoryear{{Alam} et~al.,}{{Alam}
  et~al.}{2021}]{eboss-cosmology}
{Alam} S.,  et~al., 2021, \mn@doi [\prd] {10.1103/PhysRevD.103.083533}, \href
  {https://ui.adsabs.harvard.edu/abs/2021PhRvD.103h3533A} {103, 083533}

\bibitem[\protect\citeauthoryear{Beutler, Blake, Colless, Jones, Staveley-Smith
   et~al.}{Beutler et~al.}{2012}]{Beutler:2012px}
Beutler F.,  Blake C.,  Colless M.,  Jones D.~H.,  Staveley-Smith L.,   et~al.,
  2012, \mn@doi [\mnras] {10.1111/j.1365-2966.2012.21136.x}, 423, 3430

\bibitem[\protect\citeauthoryear{{Cooke}, {Pettini}  \& {Steidel}}{{Cooke}
  et~al.}{2018}]{Cooke:2018}
{Cooke} R.~J.,  {Pettini} M.,   {Steidel} C.~C.,  2018, \mn@doi [\apj]
  {10.3847/1538-4357/aaab53}, \href
  {https://ui.adsabs.harvard.edu/abs/2018ApJ...855..102C} {855, 102}

\bibitem[\protect\citeauthoryear{Gelman \& Rubin}{Gelman \&
  Rubin}{1992}]{gelman:1992}
Gelman A.,  Rubin D.~B.,  1992, \mn@doi [Statist. Sci.]
  {10.1214/ss/1177011136}, 7, 457

\bibitem[\protect\citeauthoryear{Green}{Green}{1995}]{Green:1995}
Green P.~J.,  1995, Biometrika, 82, 711

\bibitem[\protect\citeauthoryear{Green, Godsill  \& Heikkinen}{Green
  et~al.}{2000}]{Green:2000}
Green P.~J.,  Godsill S.~J.,   Heikkinen J.,  2000. \url
  {https://api.semanticscholar.org/CorpusID:17670169}

\bibitem[\protect\citeauthoryear{Hastie \& Green}{Hastie \&
  Green}{2012}]{Hastie:2012}
Hastie D.~I.,  Green P.~J.,  2012, \mn@doi [Statistica Neerlandica]
  {https://doi.org/10.1111/j.1467-9574.2012.00516.x}, 66, 309

\bibitem[\protect\citeauthoryear{Hastings}{Hastings}{1970}]{Hastings:1970}
Hastings W.~K.,  1970, Biometrika, 57, 97

\bibitem[\protect\citeauthoryear{Hill, McDonough, Toomey  \& Alexander}{Hill
  et~al.}{2020}]{Hill:2020}
Hill J.~C.,  McDonough E.,  Toomey M.~W.,   Alexander S.,  2020, \mn@doi [Phys.
  Rev. D] {10.1103/PhysRevD.102.043507}, 102, 043507

\bibitem[\protect\citeauthoryear{Hoeting, Madigan, Raftery  \&
  Volinsky}{Hoeting et~al.}{1999}]{Hoeting:1999}
Hoeting J.~A.,  Madigan D.,  Raftery A.~E.,   Volinsky C.~T.,  1999,
  Statistical Science, 14, 382

\bibitem[\protect\citeauthoryear{Howlett, Lewis, Hall  \& Challinor}{Howlett
  et~al.}{2012}]{Howlett:2012mh}
Howlett C.,  Lewis A.,  Hall A.,   Challinor A.,  2012, \mn@doi [JCAP]
  {10.1088/1475-7516/2012/04/027}, 1204, 027

\bibitem[\protect\citeauthoryear{Karnesis, Katz, Korsakova, Gair  \&
  Stergioulas}{Karnesis et~al.}{2023}]{Karnesis:2023ras}
Karnesis N.,  Katz M.~L.,  Korsakova N.,  Gair J.~R.,   Stergioulas N.,  2023,
  {Eryn: A multi-purpose sampler for Bayesian inference} (\mn@eprint {arXiv}
  {2303.02164})

\bibitem[\protect\citeauthoryear{Karwal \& Kamionkowski}{Karwal \&
  Kamionkowski}{2016}]{Karwal:2016}
Karwal T.,  Kamionkowski M.,  2016, \mn@doi [Phys. Rev. D]
  {10.1103/PhysRevD.94.103523}, 94, 103523

\bibitem[\protect\citeauthoryear{Lewis}{Lewis}{2013}]{Lewis:2013hha}
Lewis A.,  2013, \mn@doi [Phys. Rev.] {10.1103/PhysRevD.87.103529}, D87, 103529

\bibitem[\protect\citeauthoryear{Lewis \& Bridle}{Lewis \&
  Bridle}{2002}]{Lewis:2002ah}
Lewis A.,  Bridle S.,  2002, \mn@doi [Phys. Rev.] {10.1103/PhysRevD.66.103511},
  D66, 103511

\bibitem[\protect\citeauthoryear{Lewis, Challinor  \& Lasenby}{Lewis
  et~al.}{2000}]{Lewis:1999bs}
Lewis A.,  Challinor A.,   Lasenby A.,  2000, \mn@doi [Astrophys. J.]
  {10.1086/309179}, 538, 473

\bibitem[\protect\citeauthoryear{Madigan \& Raftery}{Madigan \&
  Raftery}{1994}]{Madigan:1994}
Madigan D.,  Raftery A.~E.,  1994, Journal of the American Statistical
  Association, 89, 1535

\bibitem[\protect\citeauthoryear{Marsh \& Ferreira}{Marsh \&
  Ferreira}{2010}]{Marsh:2010}
Marsh D. J.~E.,  Ferreira P.~G.,  2010, \mn@doi [Phys. Rev. D]
  {10.1103/PhysRevD.82.103528}, 82, 103528

\bibitem[\protect\citeauthoryear{Newton \& Raftery}{Newton \&
  Raftery}{1994}]{Newton:1994}
Newton M.~A.,  Raftery A.~E.,  1994, Journal of the Royal Statistical Society.
  Series B (Methodological), 56, 3

\bibitem[\protect\citeauthoryear{Parkinson \& Liddle}{Parkinson \&
  Liddle}{2013}]{Parkinsons:2013}
Parkinson D.,  Liddle A.~R.,  2013, \mn@doi [Statistical Analysis and Data
  Mining: The ASA Data Science Journal] {https://doi.org/10.1002/sam.11179}, 6,
  3

\bibitem[\protect\citeauthoryear{{Planck Collaboration} et~al.,}{{Planck
  Collaboration} et~al.}{2020a}]{planck2018-V}
{Planck Collaboration} et~al., 2020a, \mn@doi [A\&A]
  {10.1051/0004-6361/201936386}, 641, A5

\bibitem[\protect\citeauthoryear{{Planck Collaboration} et~al.,}{{Planck
  Collaboration} et~al.}{2020b}]{planck2018-VI}
{Planck Collaboration} et~al., 2020b, \mn@doi [A\&A]
  {10.1051/0004-6361/201833910}, 641, A6

\bibitem[\protect\citeauthoryear{Poulin, Smith, Grin, Karwal  \&
  Kamionkowski}{Poulin et~al.}{2018}]{Poulin:2018}
Poulin V.,  Smith T.~L.,  Grin D.,  Karwal T.,   Kamionkowski M.,  2018,
  \mn@doi [Phys. Rev. D] {10.1103/PhysRevD.98.083525}, 98, 083525

\bibitem[\protect\citeauthoryear{Poulin, Smith, Karwal  \& Kamionkowski}{Poulin
  et~al.}{2019}]{Poulin:2019}
Poulin V.,  Smith T.~L.,  Karwal T.,   Kamionkowski M.,  2019, \mn@doi [Phys.
  Rev. Lett.] {10.1103/PhysRevLett.122.221301}, 122, 221301

\bibitem[\protect\citeauthoryear{Riess, Casertano, Yuan, Macri  \&
  Scolnic}{Riess et~al.}{2019}]{Riess:2019cxk}
Riess A.~G.,  Casertano S.,  Yuan W.,  Macri L.~M.,   Scolnic D.,  2019,
  \mn@doi [Astrophys. J.] {10.3847/1538-4357/ab1422}, 876, 85

\bibitem[\protect\citeauthoryear{Ross, Samushia, Howlett, Percival, Burden  \&
  Manera}{Ross et~al.}{2015}]{Ross:2014qpa}
Ross A.~J.,  Samushia L.,  Howlett C.,  Percival W.~J.,  Burden A.,   Manera
  M.,  2015, \mn@doi [Mon. Not. Roy. Astron. Soc.] {10.1093/mnras/stv154}, 449,
  835

\bibitem[\protect\citeauthoryear{Scolnic et~al.}{Scolnic
  et~al.}{2018}]{Scolnic:2017caz}
Scolnic D.~M.,  et~al., 2018, \mn@doi [Astrophys. J.]
  {10.3847/1538-4357/aab9bb}, 859, 101

\bibitem[\protect\citeauthoryear{Smith, Poulin  \& Amin}{Smith
  et~al.}{2020}]{Smith:2020}
Smith T.~L.,  Poulin V.,   Amin M.~A.,  2020, \mn@doi [Phys. Rev. D]
  {10.1103/PhysRevD.101.063523}, 101, 063523

\bibitem[\protect\citeauthoryear{Tierney}{Tierney}{1994}]{Tierney:1994}
Tierney L.,  1994, \mn@doi [The Annals of Statistics] {10.1214/aos/1176325750},
  22, 1701

\bibitem[\protect\citeauthoryear{Torrado \& Lewis}{Torrado \&
  Lewis}{2021}]{Torrado:2020dgo}
Torrado J.,  Lewis A.,  2021, \mn@doi [JCAP] {10.1088/1475-7516/2021/05/057},
  05, 057

\bibitem[\protect\citeauthoryear{Trotta}{Trotta}{2008}]{Trotta:2008}
Trotta R.,  2008, \mn@doi [Contemporary Physics] {10.1080/00107510802066753},
  49, 71

\bibitem[\protect\citeauthoryear{Turner}{Turner}{1983}]{Turner:1983}
Turner M.~S.,  1983, \mn@doi [Phys. Rev. D] {10.1103/PhysRevD.28.1243}, 28,
  1243

\makeatother
\end{thebibliography}




\appendix

\section{Model conditional chains weighting validation through a toy model}
\label{appendix:posterior_weighting}

As explained in Sect. \ref{Sec:BMA}, our fast model posterior computation approach (\texttt{Fast-MPC}) should achieve comparable results to a reversible jump MCMC (RJ-MCMC) routine.
To demonstrate this agreement, we present a simple example with two models. We perform BMA both with the \texttt{Fast-MPC} method and with a RJ-MCMC and compare the results from the two analyses, and we find that the methods agree to within 3.5\%.

The simple problem we choose is set up as follows: Model 1 ($M_1$) is a horizontal line $y(x) = m$ and Model 2 ($M_2$) is the line $y(x) = x+m$. The single parameter to be sampled in the MCMC, $y(0) = m$, is constrained by the data set consisting of two data points $\{(0,0),(1,0)\}$, with independent errors distributed normally with a variance of 0.5. We consider a flat model prior where each model is initially assumed to be equally likely; that is, $P(M_i)=0.5$ for i=1,2. 

We first fit each candidate model separately, running four model-conditional chains per model via standard MCMC sampler. For each chain, a preliminary set of 1000 samples is collected in order to compute the covariance of the parameters. This covariance is used as the covariance of the multivariate normal proposal distribution inside the MCMC sampler for the main conditional chains, which runs over 10,000 samples. We then use the \texttt{Fast-MPC} methodology outlined in Sect. \ref{Sec:BMA} to compute the model posteriors from the set of conditional chains. 

Next we run the RJ-MCMC chains, similarly computing the covariance for each model with 1000 initial samples and running the main chains over 10,000 samples. An RJ-MCMC sampler essentially builds a single chain from parameter samples from all the models it considers. In contrast to a classical MCMC, at each link in the chain the RJ-MCMC chooses to either accept a new point within the model it is currently in, accept a new point in another model that is more probable, or remain at its current location; in other words, the model choice is itself a parameter to be sampled. In this way, the RJ-MCMC will (after converging) choose to spend its time in each model according to the posterior probability of that model. Thus to retrieve the model posteriors from an RJ-MCMC chain, one simply counts the number of iterations the sampler spent in each model. If the number of iterations in Model 1 and Model 2 are denoted $N_1$ and $N_2$, respectively, the calculation for the model posteriors is straightforward:
\begin{equation}
    P(M_{\rm{i}}\vert\Vec{d}) = \frac{N_{\rm{i}}}{\sum_{\rm{j}=1}^K N_{\rm{j}}},
\end{equation}
where $K$ is the total number of models being compared ($K=2$ in this example). The resulting model posteriors from both BMA methods are reported in Table \ref{tab:toy_model_posteriors}.

\begin{table}
    \centering
    \begin{tabular}{lr}
    \hline\hline
    methodology                   & model posteriors     \\
    \hline
    \texttt{Fast-MPC}                      & $[ 57\%,43\% ]$      \\
    RJ-MCMC                       & $[ 58\%,42\%]$                \\
    \hline
    \end{tabular}
    \caption{Model posteriors retrieved for the toy model by \texttt{Fast-MPC} and the RJ-MCMC chains. The posteriors are reported as [$M_1$,$M_2$]. These results were retrieved using 40\% burn-in for the model conditional chains.}
    \label{tab:toy_model_posteriors}
\end{table}

To assess the agreement of the model posterior results, we ran 100 model conditional chains---50 for each model---and calculate the average model posteriors across the 50 sets of chains with the \texttt{Fast-MPC} method. We also compute the average model posteriors for 50 RJ-MCMC chains. Comparing these averages, we find that the average model posteriors agree with each other within their standard errors. The average values and their errors are reported in Table \ref{tab:toy_avg_model_posteriors}.

\begin{table}
    \centering
    \begin{tabular}{lcc}
    \hline\hline
    methodology                   & $M_1$  & $M_2$       \\
    \hline
    \texttt{Fast-MPC}                    & $(59\pm1)$\% & $(41\pm1)$\%  \\
    RJ-MCMC                     & $(57.9\pm0.3)$\% & $(42.1\pm0.3)$\%          \\
    \hline
    \end{tabular}
    \caption{Average model posteriors computed over 50 chains in each method and their standard errors, showing the agreement between the results of the \texttt{Fast-MPC} and RJ-MCMC methods for model posterior computation. The model conditional chains were all analysed with 40\% burn-in removed.}
    \label{tab:toy_avg_model_posteriors}
\end{table}

The toy model simulation successfully demonstrates that the \texttt{Fast-MPC} method for computing model posteriors is comparable in performance to the RJ-MCMC method for models that have similar posterior probabilities, and can be used to accurately compare a small number of models without the need for a more complicated RJ-MCMC infrastructure.

\section{model posterior for collapsed within-chain variance}
\label{appendix:wc-variance}
The Gelman-Rubin statistic is a valuable tool in Bayesian statistics and MCMC methods to assess convergence of multiple chains in MCMC simulations. Its primary purpose is to determine whether these chains have reached a stationary distribution and, thus, provide reliable estimates of posterior distributions.
In effect, MCMC techniques are employed to sample from complex probability distributions, particularly when analytical solutions are not available. Multiple chains are run in parallel, each starting from different initial values. The goal is to ensure that these chains converge to the same stationary distribution, which indicates that the MCMC algorithm has sufficiently explored the parameter space.
In order to assess the convergence of chains, the Gelman-Rubin statistic, expressed through the parameter R, quantifies the ratio of the \textit{between-chain} variance to the \textit{within-chain} variance. When chains have not yet converged, the between-chain variance is expected to over-estimate the converged chain variance; conversely, the within-chain variance is expected to under-estimate it.
As already introduced in Sect.~\ref{Sec:Results}, we 
adopt an analogous Gelman-Rubin statistic to assess the stability of the model posterior estimate. The within-chain variance is computed by stacking chains from individual model conditional MCMC runs. This means that we are actually computing the within-chain variance from a pair of collapsed chains from two different runs, and will refer to this quantity as the \textit{collapsed} within-chain variance.

A general prescription to compute R is as follows:
\begin{enumerate}
    \item compute the between-chain variance $\sigma^2_{\rm{BC}}$ using the mean of all chains and the variance of the chain means;
    \item calculate the within-chain variance $\sigma^2_{\rm{WC}}$;
    \item plug $\sigma^2_{\rm{BC}}$ and $\sigma^2_{\rm{WC}}$ into Eq.~\ref{eq:R} to compute R.
\end{enumerate}
For the purposes of comparing between-to within- chain variance, we treat the data $\Vec{d}$ as fixed.

As anticipated, the computation of (ii) does not come straightforwardly in \texttt{Fast-MPC}. We describe here how to compute the collapsed within-chain variance for the model posterior $P(M_{\rm{i}}\vert\Vec{d})$ defined in Eq.~\ref{eq:BMA}. The estimator $H = \hat{P}(\Vec{d}\vert M_{\rm{i}})$ introduced in Eq.~\ref{eq:BMA_like_estimator} is the harmonic mean of the conditional likelihoods $X_{\rm{t}}=P(\Vec{d}\vert M_{\rm{i}}, \Vec{\theta}_{\rm{i}}^{(\rm{t})})$ for the model $M_{\rm{i}}$. 

It is sometimes more convenient to consider the reciprocal of $H$:
\begin{equation}
    H^{-1} = \frac{1}{N}\sum_{t=1}^{N}\frac{1}{X_{\rm{t}}},
\end{equation}
where we kept the notation introduced in the main text. Let us therefore define the quantities
\begin{align}
    \mu_{\rm{R}} &= \rm{E}(X_t^{-1}) \\
    \sigma^2_{\rm{R}} &= \rm{Var}(X_t^{-1})
\end{align}
where R stands for \textit{reciprocal}.

Using the central limit theorem, provided that in our case $X_{\rm{t}}^{-1}$ has finite variance 
we can demonstrate that 
\begin{equation}
    \sqrt{N}(H^{-1} - \mu_{\rm{R}})\rightarrow N(0,\sigma^2_{\rm{R}})
\end{equation}
and using the delta method for a function $g(x)=x^{-1}$ we obtain the useful result:
\begin{equation}
    \sqrt{N}(H-\mu_{\rm{R}}^{-1})\rightarrow N\left(0,\frac{\rm{Var(X^{-1}_{\rm{t}})}}{\rm{E}(X_{\rm{t}}^{-1})^4}\right),
\end{equation}
which in our case gives
\begin{equation}
    \sigma^2(H) = \frac{\sigma^2_{\rm{R}}}{N\mu_{\rm{R}}^4}.
    \label{eq:var_model}
\end{equation}
In our analysis, $H=H(M_{\rm{i}})$ for $\rm{i}=1,2$, and we want to estimate the variance of $P(M_{\rm{i}}\vert\Vec{d})$. Without loss of generality, consider $i=1$:
\begin{equation}
    \rm{Var}(\hat{P}(M_{\rm{1}}\vert\Vec{d})) \approx \rm{Var}\left(1-\frac{P(M_2)\hat{P}(\Vec{d}\vert M_{\rm{2}})}{P(M_1)\hat{P}(\Vec{d}\vert M_{\rm{1}})}\right),
    \label{eq:var_PM}
\end{equation}
where we used the first order Taylor expansion of $1/x\approx 1 - (x-1)$ applied to the definition of $P(M_{\rm{i}}\vert\Vec{d})$ given in Eq.~\ref{eq:BMA}.
The first order Taylor expansion of the distribution $f(M_1,M_2) = 1 - \frac{\hat{P}(\Vec{d}\vert M_{\rm{2}})}{\hat{P}(\Vec{d}\vert M_{\rm{1}})}$ around 
\begin{equation}
   \Vec{\theta_0}= \left(\mu_1=\mu(\hat{P}(\Vec{d}\vert M_1)),\mu_2=\mu(\hat{P}(\Vec{d}\vert M_2))\right) 
\end{equation} 
provides an estimator of Eq.~\ref{eq:var_PM}:
\begin{equation}
\begin{split}
    \rm{Var}(\hat{P}(M_{\rm{1}}&\vert\Vec{d})) \approx \frac{1}{\mu_2^2}P^2(M_1)\sigma^2(\hat{P}(\Vec{d}\vert M_1))-\\
    &2P(M_1)P(M_2)\frac{\mu_1}{\mu_2^3}\rm{Cov}(\hat{P}(\Vec{d}\vert M_1),\hat{P}(\Vec{d}\vert M_2))+\\
    &\frac{\mu_1^2}{\mu_2^4}P^2(M_2)\sigma^2(\hat{P}(\Vec{d}\vert M_2))) =\\
    &\frac{\mu_1^2}{\mu_2^2}\left[ \frac{P^2(M_1)\sigma^2(\hat{P}(\Vec{d}\vert M_1))}{\mu_1^2} + \frac{P^2(M_2)\sigma^2(\hat{P}(\Vec{d}\vert M_2))}{\mu_2^2} \right],
\end{split}
\label{eq:wv_estimate}
\end{equation}
where we neglected the covariance term by means of the independence of estimator $\hat{P}(\Vec{d}\vert M_{\rm{i}})$ for different i. That is, treating $\Vec{d}$ as fixed, the covariance term is equal to zero by the independence of model-conditional chains whose samples are separately used to compute $\hat{P}(\Vec{d}\vert M_{\rm{i}})$, $i=1,2$. Note that $\Vec{d}$ is treated as fixed (i.e., not random) in the calculation of between-chain variance because all chains use the same data. In order for the within-chain variance to be comparable to the between-chain variance, it too should treat $\Vec{d}$ as fixed. This means the randomness in $\hat{P}(\Vec{d}\vert M_i)$ is attributed to the parameters randomly sampled in different chains. Since each chains is run independently, the samples from different chains are also independent.

Eq.~\ref{eq:wv_estimate} can be conveniently evaluated from the quantities $\mu_i$ and $\sigma^2(\Vec{d}\vert P(M_{\rm{i}}))$ defined by Eq.~\ref{eq:BMA_like_estimator} and Eq.~\ref{eq:var_model} respectively. Using this formula we can determine the within-chain variance, and compare against the between-chain variance to assess estimation stability for the \texttt{Fast-MPC} method.


\bsp	
\label{lastpage}
\end{document}